\documentclass[a4paper,11pt]{article}
\usepackage{jcappub} 

\usepackage{graphicx}
\usepackage{soul}
\usepackage{dcolumn}
\usepackage{color}
\usepackage{xcolor}
\usepackage{enumitem}
\usepackage{cancel}
\usepackage{mathtools}
\usepackage{hyperref}
\usepackage{orcidlink}
\usepackage{textgreek}

\newcommand{\dd}{{\rm d}}

\newcommand{\expect}[1]{\left\langle #1 \right\rangle}
\newcommand{\DM}{{\rm DM}}
\newcommand{\pfrac}[2]{\left(\frac{#1}{#2}\right)}

\arxivnumber{2311.07648} 
\title{Coherent Self-Interactions of Dark Matter in the Bullet Cluster}
\date{\today}

\author[a,b]{Zachary Bogorad\,\orcidlink{0000-0001-9913-6474}\,}
\author[a,b]{Peter W.~Graham\,\orcidlink{0000-0002-1600-1601}\,}
\author[a,c]{Harikrishnan Ramani\,\orcidlink{0000-0003-3804-8598}\,}

\affiliation[a]{Stanford Institute for Theoretical Physics, Department of Physics, Stanford University, Stanford, CA 94305, USA}
\affiliation[b]{Kavli Institute for Particle Astrophysics \& Cosmology, Department of Physics, Stanford University, Stanford, CA 94305, USA}
\affiliation[c]{Department of Physics and Astronomy, University of Delaware, Newark, DE 19716, USA}

\emailAdd{zbogorad@stanford.edu}

\abstract{Many models of dark matter include self-interactions beyond gravity. A variety of astrophysical observations have previously been used to place limits on the strength of such self-interactions. However, previous works have generally focused either on short-range interactions resulting in individual dark matter particles scattering from one another, or on effectively infinite-range interactions which sum over entire dark matter halos. In this work, we focus on the intermediate regime: forces with range much larger than dark matter particles' inter-particle spacing, but still shorter than the length scales of known halos. We show that gradients in the dark matter density of such halos would still lead to observable effects. We focus primarily on effects in the Bullet Cluster, where finite-range forces would lead either to a modification of the collision velocity of the cluster or to a separation of the dark matter and the galaxies of each cluster after the collision. We also consider constraints from the binding of ultrafaint dwarf galaxy halos, and from gravitational lensing of the Abell 370 cluster. Taken together, these observations allow us to set the strongest constraints on dark matter self-interactions over many orders of magnitude in range below $\sim10$ kpc, surpassing existing limits by orders of magnitude throughout.}

\begin{document}
\maketitle
\flushbottom

\section{Introduction}\label{sec:Introduction}

Although the presence of dark matter in the universe has been firmly established by a variety of experiments, its nature and properties remain essentially unknown. In particular, while dark matter's presence is inferred from its gravitational effects, it remains undetermined whether dark matter particles have other interactions with the Standard Model or with each other.

The presence of possible self-interactions of dark matter is interesting for a variety of reasons. First, in the absence of large interactions between a dark sector and the Standard Model, self-interactions might become the first discernible feature of a dark sector. Second, certain short-scale properties of galaxies appear to be in tension with the results of simulations in the absence of dark matter self-interactions, including the core-cusp problem (see, for example, Refs. \cite{CoreCusp1, CoreCusp2, CoreCusp3}) and the missing satellites problem (Refs. \cite{MissingSatellites1, MissingSatellites2, MissingSatellites3}); see also Ref. \cite{SIDMDiscrepancyReview} for a review of various related discrepancies. While some recent work has suggested that these individual discrepancies can be explained without the need for dark matter self-interactions, it remains unclear whether existing astrophysical observations can be fully explained by dark matter that interacts only through gravity; see, for example, Refs. \cite{Explanations1, Explanations2, Explanations3, Explanations4}.
Self-interactions are also motivated by some models of dark energy. In particular, quintessence models can lead to new long-ranged interactions between dark matter particles \cite{QuintessenceDMSI, DMSIStructureFormation}.

Self-interactions of dark matter have been probed via a variety of astrophysical observations, leading to a range of constraints on self-interaction couplings as a function of the force's range (and, in some cases, on other properties of the dark matter). However, most such constraints treat these interactions in one of two limits: a short-range limit in which dark matter particles interact with one another only individually; and a long-range limit in which the force is summed over the entirety of the considered dark matter halo. Intermediate ranges, in which collective interactions are significant but the force range remains smaller than the halo size, have rarely been considered.  A simplistic interpolation between the short- and long-range regimes would assume that forces which are exponentially suppressed beyond their range would lead to exponentially suppressed effects in halos larger than that range.

They key point of this work is that this is not the case: forces with ranges that are large compared to the interparticle spacing (or, if dark matter is strongly clumped, the inter-clump spacing) but smaller than the size of the containing dark matter halo can still have large collective effects, due to the presence of density gradients within such halos. This will allow us to set substantially strengthened constraints on dark matter self-interactions with ranges below approximately $\sim10~{\rm kpc}$, surpassing existing constraints on self-interactions by many orders of magnitude in strength over many orders of magnitude in range.

The rest of this work is organized as follows: In Section \ref{sec:CollisionVelocity}, we present a constraint on long-ranged attractive forces based on observations of the Bullet Cluster. (Repulsive forces are dealt with in Section.~\ref{sec:OtherConstraints}). As we will show, the strongest constraint in the attractive, long-ranged regime arises not from the scattering of dark matter particles between the two clusters, but rather from a combination of the observed collision velocity of the clusters' gases and the coincidence of the location of the dark matter and the galaxies after the collision. We emphasize, however, that the behavior we obtain in this section for the Bullet Cluster is in fact fairly general, and can be used to extend a variety of existing constraints on long-ranged dark matter self-interactions to finite ranges. In Section \ref{sec:OtherConstraints} we will consider two such constraints: the self-binding of ultrafaint dwarf galaxy's halos to constrain repulsive interactions, and the agreement of mass measurements of the Abell 370 cluster using gravitational lensing and using its galaxies' velocity dispersion, applicable to both attractive and repulsive forces. Notably, the latter will also allow us to illustrate how the existence of substructure within halos can affect our conclusions, potentially precluding some bounds from being extended to shorter ranges. Finally, in Section \ref{sec:Conclusion}, we summarize our results.

\section{Self-Interaction Constraints from the Bullet Cluster}\label{sec:CollisionVelocity}

The Bullet Cluster (1E 0657–56) is a pair of galaxy clusters whose mass and velocity distributions indicate that they recently collided \cite{BCFirstObservation, BCFollowUpObservation, BCPassage}. In particular, while the mass distributions and luminous components of both clusters are now well-separated \cite{BCFirstLensing, BCLensing2, BCLensing3}, the intracluster gas components of both clusters have now merged into a largely continuous distribution between the clusters' mass distributions, with prominent shock waves arising from their supersonic collision \cite{BCShockMeasurement1, BCShockMeasurement2}. The fact that this did not occur for the bulk of the clusters' mass indicates the presence of a large mass of sufficiently non-collisional matter in both clusters---dark matter---and sets limits on how much this dark matter can interact with itself in order to pass through the other cluster largely unaffected. Estimating this effect in the presence of a hard scattering interaction between dark matter particles leads to a largely model-agnostic constraint on the momentum transfer cross-section $\sigma_T$ for dark matter particles of mass $m$ to scatter with one another (see, for example, Refs. \cite{BCFirstDMSILimits, LDMReviewKLZ, BCDMSILimits2, BCDMSILimits3}):
\begin{align}
    \frac{\sigma_T}{m} \lesssim 1~\frac{{\rm cm}^2}{\rm g},
\end{align}
up to order-one factors that differ between sources. 

The momentum transfer cross-section $\sigma_T$ is straightforward to evaluate for short-range interactions, where only collisions between individual particles matter. In this work we are interested in the opposite regime, where the new force's range is large and collective effects start becoming important. For concreteness, let us focus on Yukawa forces, such that the potential $V$ between two particles of masses $m_{1,2}$ separated by a distance $r$ is given by
\begin{align}
    V &= -\alpha G \frac{m_1 m_2}{r}e^{-r/\lambda}, \label{eq:YukawaPointPotential}
\end{align}
with $\alpha$ the new force's strength relative to gravity, $\lambda$ its range, and $G$ Newton's constant. For force ranges much smaller than the inter-particle spacing, the impact parameters that contribute to scattering are also much smaller than the inter-particle spacing, and hence collective effects are not important. However, when the force range $\lambda$ exceeds the inter-particle spacing, each particle scatters coherently off a whole mass distribution which greatly enhances the interaction. 

We will consider two main classes of interactions: purely attractive, and purely repulsive. Note that the former can mostly simply arise from the exchange of a scalar mediator, while the latter can arise from a vector mediator, assuming the minimal situation where all dark matter particles have the same charge. We describe these models in more detail, and discuss the fine-tuning required for the parameter space we consider as well as these models' behavior in finite-density dark matter backgrounds, in App. \ref{app:FiniteDensityEffects}.

An extremely conservative bound on these interactions can be still be set by restricting ourselves to individual scatters with impact parameter much less than the inter-particle spacing, i.e by setting the maximum impact parameter $b_{\rm max}$  to obey $b_{\rm max} \ll  n^{-1/3}$, where $n$ is the dark matter number density. If we also require that $b_{\rm max}\ll \lambda$, we can approximate the Yukawa interaction as a Coulomb interaction. In this limit, $\sigma_T$ can be evaluated to be \cite{PBHConstraintsUFD}
\begin{align}
\sigma_T&=\frac{8 \pi  \kappa^2}{m_{\rm DM}^2 v_{\rm rel}^4}
\log \left(\frac{2\kappa^2+b_{\rm max}^2 m_{\rm DM}^2
   v_{\rm rel}^4}{2\kappa^2+b_{\rm min}^2 m_{\rm DM}^2
   v_{\rm rel}^4}
   \right)\\
   \nonumber    
   &=\frac{16\pi \alpha^2 G^2 m_{\rm DM}^2}{v_{\rm rel}^4} \ln \Lambda.
\end{align}
Here $m_{\rm DM}$ is the mass of individual dark matter particles, $v_{\rm rel}$ is the relative velocity of the collisions, $b_{\rm min}$ is the minimum impact parameter set by the typical distance of closest approach of a particle, $\kappa=\alpha G m_{\rm DM}^2$ and $\log \Lambda$ is the Coulomb logarithm, with
\begin{align}
    \Lambda^2 = \frac{2\kappa^2+b_{\rm max}^2 m_{\rm DM}^2 v_{\rm rel}^4}{2\kappa^2+b_{\rm min}^2 m_{\rm DM}^2 v_{\rm rel}^4} .
\end{align}
The bound resulting from requiring that this momentum-transfer cross-section be less than $1~{\rm cm}^2/{\rm g}$ is plotted for two example dark matter masses in Figures \ref{fig:AttractivePlot} and \ref{fig:RepulsivePlot} (labelled ``BC-Incoherent''), using
\begin{align}
    &\pi b_{\rm min}^2 (2r_A) \frac{\rho_A}{m_{\rm DM}} = 1 \nonumber \\
    &b_{\rm max} = \min\left( \lambda,~0.1 \left(\frac{\rho_A}{m_{\rm DM}}\right)^{-1/3} \right)
\end{align}
to estimate the Coulomb logarithm, with $\rho_A$ and $r_A$ the central density and the scale radius of the larger cluster in the Bullet Cluster (discussed in more detail below Eq. \eqref{eq:NFWProfile}). Here, $b_{\rm min}$ is the smallest distance at which a given dark matter particle expects to encounter one other particle during its passage, and $b_{\rm max}$ is set by either the force's range or the inter-particle spacing (including a conservative safety factor of $0.1$); whichever is smaller.

However, it is immediately clear that the above constraint is too weak, and inclusion of coherent effects should lead to much stronger limits on the coupling $\alpha$.

One such effect is the change in the relative velocity of the clusters at the time of collision due to extra acceleration felt by dark matter. This can lead to one of two observable effects, depending on whether the dark matter and the Standard Model components of each cluster separate or remain bound together until the collision. Some of these effects have been considered before in the context of forces with range much larger than a Mpc; see, for example, Refs. \cite{VariousLRDMSIConstraints, BCCollisionNotUnexpected}.

Our focus, however, will be on forces with ranges shorter than the length scales of the clusters in the Bullet Cluster, such that their behavior differs significantly from that of infinite-range forces of the same strength. This regime has not, to our knowledge, been evaluated previously. We will specifically focus on Yukawa forces (Eq. \eqref{eq:YukawaPointPotential}), although most of our conclusions can be straightforwardly extended to other potentials. Crucially, despite the forces we consider falling off exponentially at distances beyond their range, we show below that the resulting bounds on interaction strength are suppressed only as the square of the ratio of cluster length scale to force range, not exponentially. Note that, in this section, we will focus our discussion on attractive forces: repulsive interactions of dark matter can be constrained more strongly at all ranges by the straightforward arguments of Section \ref{sub:HaloBinding}.

Measurements of dark matter halo profiles via gravitational lensing are difficult, so the precise shapes and other properties of the two clusters within the Bullet Cluster remain somewhat uncertain (see, for example, Refs. \cite{BCFirstDMSILimits, BCEvidenceForDMSI, BCCollisionNotUnexpected, LensingUncertainties}). In this work, we will assume the same profiles used in Ref. \cite{BCCollisionNotUnexpected}, as our primary purpose is to point out the finite-range behavior of an effect it discusses, but we note that different profile assumptions may change our results by order-one factors; see Sec. \ref{sub:DensityProfile}. In particular, we will assume that both clusters---the larger (hereafter ``A'') cluster, and the smaller ``bullet" (hereafter ``B'') cluster---possess an NFW density profile given by
\begin{align}
    \rho(r)=\frac{M_{200}}{4\pi (r^*)^3 \zeta(c)}\frac{1}{(r/r^*)(1+r/r^*)^2} \label{eq:NFWProfile}
\end{align}
where $M_{200}$ is the virial mass of the cluster, $\zeta(c)=\log(1+c)-c/(1+c)$, $r^*=r_{200}/c$, is the scale radius, $r_{200}$ is the virial radius, and $c$ is the concentration parameter. We use $M_{200,A} = 1.5\times10^{15}~M_\odot$, with radius $r_{200,A} = 2136~\text{kpc}$ and concentration $c_{200,A} = 1.94$, and $M_{200,B} = 1.5\times10^{14}~M_\odot$ with $r_{200,B} = 995~\text{kpc}$ and $c_{200,B} = 7.12$. It should be understood, however, that there are substantial uncertainties in the properties of the Bullet Cluster (and especially of its dark matter halos), which can significantly affect its inferred dynamics \cite{BCSeparationUncertainty, ClusterDMUncertainties}. Our calculations in this work are therefore limited to order-of-magnitude estimates, and we set correspondingly conservative constraints, excluding only new forces that would cause large changes in the Bullet Cluster collision.

Having stated these assumptions, we next turn to computing the relative velocity changes of the Bullet Cluster components, and the observational consequences of these changes.

\subsection{Separation of the Dark Matter and the Standard Model}\label{sub:Separation}

We first consider the possibility of the dark matter separating from the Standard Model components (both gas and stars) due to the presence of an extra force felt by the dark matter alone. Note that this is distinct from what is observed: the gas alone separates and the stars in galaxies and dark matter remain together through the collision. In particular, the present-day positions of both the dark matter and the stars of the two clusters are observable via gravitational lensing and optical measurements. The separation between the centroids of these two components for the smaller of the two clusters has been observed to be $25 \pm 29~{\rm kpc}$ \cite{BCDMSILimits2, BCLensing2}, although estimating this is difficult and later analysis indicates that this uncertainty should be closer to $40$ kpc \cite{BCSeparationUncertainty}. Regardless, this is significantly smaller than the size of either cluster, which immediately precludes situations in which the dark matter and the Standard Model gas within it become separated during approach. We convert this into a limit on the dark matter coupling $\alpha$ next.

Let us begin by considering a  test dark matter particle of mass $m$ in cluster B falling into cluster A. We use the dark matter profile described above for the A cluster. To minimize dependence on the clusters' halo profiles, we can limit the additional acceleration it receives from the new force to its value at A's scale radius, thus reducing the impact of any uncertainty in the central density. This will give us a conservative lower bound on the dark matter-Standard Model separation. The acceleration due to gravity alone on the dark matter in the B cluster and the Standard Model in the B cluster are identical because of the equivalence principle and hence will not be considered for evaluating the separation. The net Yukawa potential obtained by integrating over the new force's Yukawa potential, Eq.~\eqref{eq:YukawaPointPotential}, over the A cluster profile gives:
\begin{align}
    \frac{V_{\rm new}(\mathbf{r_A^*})}{m} &= -\alpha G \int \dd^3\mathbf{r} \frac{\rho_A(\mathbf{r})}{|\mathbf{r}-\mathbf{r_A^*}|}e^{-|\mathbf{r}-\mathbf{r_A^*}|/\lambda} \label{eq:YukawaPotentialIntegral}
\end{align}
where $\mathbf{r_A^*}$ is any point at the scale radius of the A cluster. For $\lambda \ll r_A^*$, the additional acceleration seen by the dark matter in cluster B while at radii $r' \sim r_A^*$ is, from Eq. \eqref{eq:YukawaPotentialIntegral}, of order
\begin{align}\begin{split}
    a_{\rm DM} &= \frac{\dd}{\dd r'}\left( -\alpha G \int \dd^3\mathbf{r} \frac{\rho_A(\mathbf{r})}{|\mathbf{r}-\mathbf{r'}|}e^{-|\mathbf{r}-\mathbf{r'}|/\lambda} \right) \\
    &\sim \frac{\dd}{\dd r'}\left( -\alpha G \frac{\lambda^3}{\lambda} \rho_A(r') \right) \\
    &\sim -\alpha G \lambda^2 \frac{\rho_A(r_A^*)} {r_A^*} \label{eq:NewAccelerationDMShort}
\end{split}\end{align}
where we have taken advantage of the fact that the density of A varies over the long length scale $r_A^*$, whereas the Yukawa potential exponentially suppresses contributions from farther than the much shorter length scale $\lambda \ll r_A^*$. (There may of course be shorter-length scale variations in $r_A^*$ due to substructure, but, as we discuss below, this should not affect our conclusions so long as the dark matter halo also has a non-negligible smooth component.) For larger ranges ($\lambda > r_A^*$), the new force behaves simply like gravity: $a_{\rm DM} \sim \alpha G\rho_A(r_A^*) r^*_A$; see Eq. \eqref{eq:GravityAccelerationScaleRadius} below. We can conveniently combine these limiting regimes as
\begin{align}\begin{split}
    a_{\rm DM} &\sim -\alpha G \rho_A(r_A^*) r_A^* \left(1 + \frac{\lambda^2}{(r_A^*)^2} \right) \label{eq:NewAccelerationDM}.
\end{split}\end{align}

Note that, as advertised, the acceleration scales as a power law in $\lambda$, as seen in the last line of Eq.~\eqref{eq:NewAccelerationDM}, despite the first line containing the Yukawa exponent. This scaling can also be obtained from the following simple scaling argument: For any power-law profile with length scale $r_A^* \gg \lambda$, there are masses of order $\rho_A(r_A^*) \lambda^3$ within a distance of order $\lambda$ both inward ($M_{\rm in}$) and outward ($M_{\rm out}$) of the chosen point, but the inward direction is more dense by a fraction of order $\lambda/r_A^*$, such that $M_{\rm in}-M_{\rm out} \sim \rho_A(r_A^*) \lambda^3 (\lambda/r_A^*)$. Thus, roughly, the two sides provide acceleration 
\begin{align}\begin{split}
    a_{\rm DM} &\sim \frac{\alpha G M_{\rm out}}{\lambda^2}-\frac{\alpha G M_{\rm in}}{\lambda^2} \\
    &\sim -\frac{\alpha G }{\lambda^2}\left(\rho_A(r_A^*) \lambda^3\right) \frac{\lambda}{r_A^*} \\
    &\sim -\alpha G \frac{\rho_A(r_A^*) \lambda^2}{r_A^*}. \label{eq:NewAccelerationScaling}
\end{split}\end{align}
This reproduces the power-law behavior of Eq. \eqref{eq:NewAccelerationDM}.

The next question is whether the stars are dragged along with the dark matter as it is accelerated. This happens purely gravitationally. There is a maximum acceleration of dark matter that the stars can keep up with. Conservatively, we can estimate this acceleration by evaluating the maximum acceleration of stars in the B cluster as a function of separation from its center. In the absence of a length scale other than the scale radius, this is given, parametrically, by
\begin{align}\begin{split}
    a_g &\sim -G \frac{\rho_B(r^*_B) (r^*_B)^3}{(r^*_B)^2} \sim -G \rho_A(r^*_A) r^*_A \left[ \frac{\rho_B(r^*_B) r^*_B}{\rho_A(r^*_A) r^*_A} \right] \\
    &\sim -G \rho_A(r^*_A) r^*_A \label{eq:GravityAccelerationScaleRadius}
\end{split}\end{align}
where we have taken advantage of the bracketed ratio being order-one (see Eq. \eqref{eq:NFWProfile} and the values below it). While some profiles may have accelerations larger than this estimate by order-one factors, large changes to this expression are possible only for clusters that are parametrically more cuspy than e.g. an NFW profile. In any case, we will see below that our constraints are independent of this value. 

If $a_{\rm DM}$ in  Eq.~\eqref{eq:NewAccelerationDM} exceeds $a_g$ in Eq.~\eqref{eq:GravityAccelerationScaleRadius}, then the dark matter matter separates. The net separation between stars and dark matter $\Delta x$, over the entire bullet cluster event duration $t_{\rm bc}\approx r_A^*/(v_A+v_B)$, is then given by
\begin{align}\begin{split}
    \Delta x &\sim \left(a_{\rm DM}-a_{g}\right)t_{\rm bc}^2 \\
    &\sim a_{\rm DM}\left(1-\frac{a_g}{a_{\rm DM}}\right)\left(\frac{r_A^*}{v_A+v_B}\right)^2 \\
    &\sim r_A^* \left(1-\frac{a_g}{a_{\rm DM}}\right)
\end{split}\end{align} 
In the last line we have approximated the velocities via $v^2/r_A^* \sim a_{\rm DM}$. Thus, even if $a_{\rm DM}$ only exceeds $a_g$ by $\mathcal{O}(1)$, the separations are $\mathcal{O}\left(r_A^*\right)$, in severe tension with observation. Then, requiring that $a_{\rm DM}$ not exceed $a_g$ in Eqs.~\eqref{eq:NewAccelerationDM} and ~\eqref{eq:GravityAccelerationScaleRadius}, we rule out the couplings
\begin{align}
    \alpha \gtrsim 1+\left(\frac{r_A^*}{\lambda}\right)^2 
    \label{eq:separate}
\end{align}
where the $1$ in front is required for the correct asymptote at range larger than the clusters; see the discussion below Eq. \eqref{eq:NewAccelerationDM}.

\subsection{Modification of the Gas Collision Velocity}\label{sub:CollisionVelocity}

If the dark matter and gas components of the clusters remain together as the clusters approach one another, the relative velocity of the dark matter is (indirectly) observable thanks to the visible shock fronts in the colliding gas, whose shapes are a function of the Mach number of that collision. Historically, a disagreement between this observed velocity and the collision velocity expected from gravitational infall between clusters of the observed masses has been cited as evidence for long-ranged interactions of dark matter \cite{BCEvidenceForDMSI}, although improved observations of the clusters combined with detailed simulations of the collision dynamics have brought these values into reasonable agreement assuming gravitational interactions alone \cite{BCCollisionNotUnexpected}. As Ref. \cite{BCCollisionNotUnexpected} notes, this agreement can thus be reinterpreted as a bound on new long-ranged interactions, which could modify the collision velocity. This is precisely the bound that we will be evaluating in this section. For the first time, we also extend this bound to force ranges much shorter than the typical length scales of the Bullet Cluster.

To evaluate this bound, let us now consider the case when $a_g$ exceeds $a_{\rm DM}$, such that the dark matter and Standard Model arrive together. We begin by considering the fall of a single particle from rest at infinity into a cluster, in the limit of negligible friction due to scattering with other individual particles, such that the velocity of the particle at any point can be evaluated simply from the potential relative to infinity seen by the particle at that point. Requiring that a new force not change the collision velocity of that particle with the cluster by more than order unity is thus equivalent to requiring that the potential at the collision point not be changed by more than order unity, i.e.
\begin{align}
    \frac{ \left| V_{\rm new}(\mathbf{r_A^*}) \right| }{m} \lesssim \frac{ \left| V_{\rm g}(\mathbf{r_A^*}) \right| }{m} = G \int \dd^3\mathbf{r} \frac{\rho_A(\mathbf{r})}{|\mathbf{r}-\mathbf{r_A^*}|}. \label{eq:ScaleRadiusPotentialComparison}
\end{align}
where $V_{\rm new}(\mathbf{r_A^*})$ is given in Eq.~\eqref{eq:YukawaPotentialIntegral}. Before we proceed further, let us obtain an order-of-magnitude estimate of Eq.~\eqref{eq:YukawaPotentialIntegral} in order to understand its scaling behavior. A simple approximation for the potential per unit mass at radius $r_A^*$ can obtained from arguments similar to Eq.~\eqref{eq:NewAccelerationScaling} to obtain
\begin{align}
    \frac{V_{\rm new}^*}{m} \sim a^* r_A^* \sim \alpha G \rho_A(r_A^*)\left(1+ \left(\frac{\lambda}{r_A^*}\right)^2\right), \label{eq:NewPotentialScaleRadiusGeneric}
\end{align}
Requiring this to be greater than $V_g \sim G \rho_A(r_A^*) \left(r_A^*\right)^2$ gives   
\begin{align}
    \alpha \gtrsim 1+\left(\frac{r_A^*}{\lambda}\right)^2 
\label{eq:together}
\end{align}
which is very similar to Eq.~\eqref{eq:separate}. 

\begin{figure}[t]
    \centering
    \includegraphics[width=0.7\linewidth]{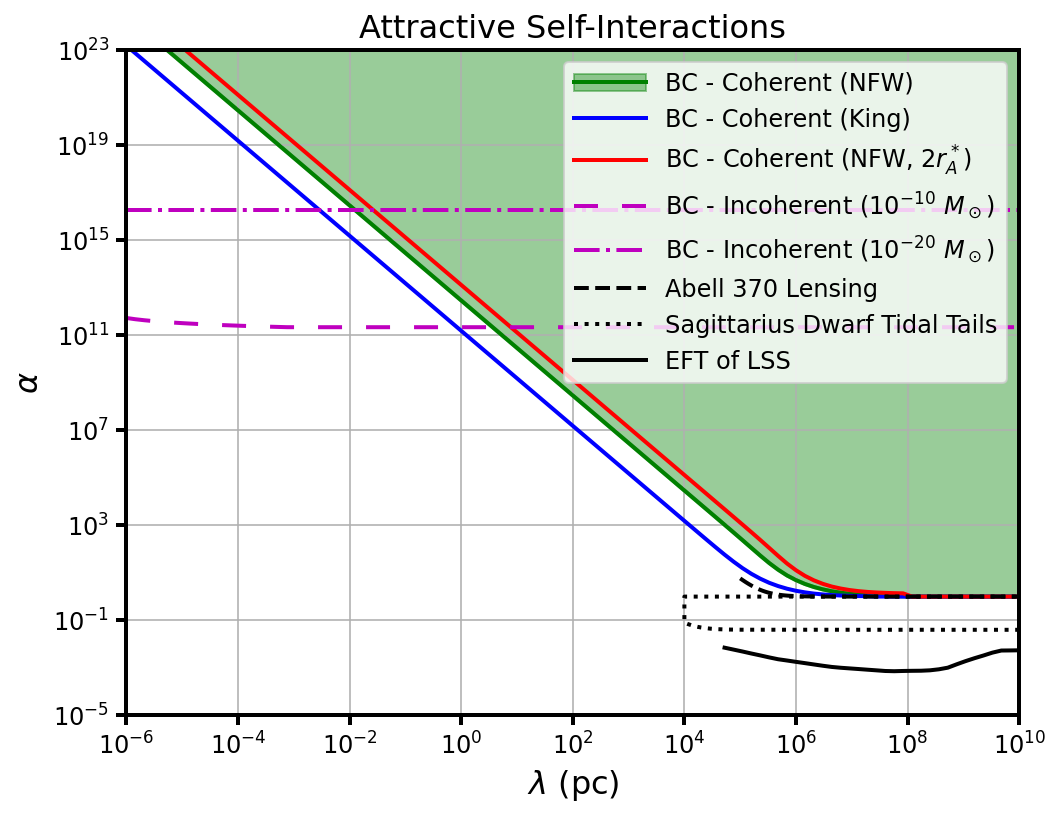}
    \caption{Our constraints on attractive self-interactions of dark matter beyond gravity from coherent interactions in the Bullet Cluster (see Section \ref{sec:CollisionVelocity}) and from mass measurements of the Abell 370 cluster (see Section \ref{sub:GravitationalLensing}), and, for comparison, previous constraints from the tidal streams of the Sagittarius dwarf galaxy \cite{TidalTails1, TidalTails2, TidalTailsUpperBound}, from incoherent scattering in the Bullet Cluster \cite{BCFirstDMSILimits, LDMReviewKLZ, BCDMSILimits2, BCDMSILimits3} at two assumed dark matter particle masses (the largest mass at which promordial black holes can make up all of dark matter, $\sim 10^{-10}~M_\odot$ \cite{PBHDMReview}, as well as $\sim 10^{-20}~M_\odot$ to illustrate the mass dependence), and from cosmological effects of a background field of the mediator \cite{DarkForceBackgroundEffects1, DarkForceBackgroundEffects2, DarkForceBackgroundEffects3}, which have recently been tentatively extended to $\lambda \gtrsim$ 40 kpc. As a rough estimate of the uncertainty on our bound, the collision velocity constraint is shown assuming both NFW and King profiles, as well as for an NFW profile but conservatively assuming that the gas collides at two times the scale radius rather than at the scale radius of the A cluster. The shortest ranges at which we plot the lensing and tidal tails constraints are approximate; see the discussion in Section \ref{sec:OtherConstraints}. Note also that the upper edge of the parameter space excluded by the Sagittarius Dwarf is presently unknown; Ref. \cite{TidalTailsUpperBound} has shown only that $\alpha=1$ is consistent with observation.}
    \label{fig:AttractivePlot}
\end{figure}

While the order-of-magnitude estimates match for the two observations, numerical evaluation shows that the constraint obtained from the non-observation of the separation of dark matter and the stellar population is stronger. This occurs primarily from the fact that even a small mismatch between $a_{\rm DM}$ and $a_g$ results in a separation of $\mathcal{O}\left(r_A^*\right)$, whereas this separation is constrained to be a tiny fraction of $r_A^*$. 
In the spirit of staying conservative---e.g. with respect to additional interactions between dark matter and the Standard Model or to effects of substructure which might suppress the observation in Section \ref{sub:Separation}---we use the weaker bound obtained from numerically evaluating the bound in Eq.~\eqref{eq:ScaleRadiusPotentialComparison}. This conservative constraint is plotted in Fig. \ref{fig:AttractivePlot} as ``BC - Coherent (NFW)''. (The other ``BC - Coherent'' lines are modifications to this bound discussed below.)

There are several complications to this analytical description of the collision in the presence of a new force. We discuss them over the next several subsections.

\subsection{Halo Density Profiles}\label{sub:DensityProfile}

We default to NFW profiles for both clusters in this work due to their widespread use, but marginally better lensing fits are, in fact, obtained using King profiles \cite{BCEvidenceForDMSI}. This leads to only small changes in the result of numerically evaluating the bound in Eq. \eqref{eq:ScaleRadiusPotentialComparison}, as seen in Fig. \ref{fig:AttractivePlot} (see the ``BC - Coherent (King)'' lines). The minimal dependence on profile choice is explained by the scaling arguments that lead to Eq. \eqref{eq:NewAccelerationScaling}: any power-law profile with length scale $r_A^* \gg \lambda$ will lead to accelerations and potentials which scale as $\left(\lambda/r_A^*\right)^2$ for small $\lambda$. 

\subsection{Substructure}\label{sub:Substructure}

A related caveat to our estimate above is that it assumes that all dark matter is smoothly distributed in the shape of this profile. In fact, this is certainly not the case: the presence of galaxies within the two clusters indicates that at least some fraction of the dark matter in each cluster is located in dense galactic halos, and might therefore not interact with the dark matter in galaxy halos of the other cluster unless the two galactic halos happen to pass within a distance of order $\lambda$ of one another. So long as at least an order-one fraction of the dark matter in the A cluster remains outside its galaxies, however---as is expected from N-body simulations of structure formation (see, for example, Ref. \cite{StructureFormationSimulation} for a review)---this should not affect our results by more than order one. Put equivalently, if the dark matter distribution in cluster A can be described as the sum of a smooth component of mass $M_{\rm smooth}$ and a clumped component (e.g. from galaxies) of mass $M_{\rm clumpy}$, our discussion so far will hold unchanged for the smooth component. (The clumpy component may have no effect, if halos never approach one another closely enough, or may also contribute if they do.) Our results are thus not changed (beyond order-one factors) unless $M_{\rm smooth} \ll M_{\rm clumpy}$, which is not expected.

More generally, our results hold only when the dark matter in the two clusters is not fully sequestered in clumps separated by length scales longer than the force range $\lambda$. Such large separations could result in dark matter clumps of cluster B passing through cluster A without ever passing within $\lambda$ of another clump, thus never being affected by the new force. While this scenario appears unlikely and is not supported by structure formation simulations \cite{StructureFormationSimulation}, it may be possible to construct models in which our constraint is evaded in this way. We emphasize, however, that while the presence of a new finite-range attractive force affecting dark matter would certainly increase the clustering of dark matter at shorter length scales, our bound would not be modified significantly unless the smoothly-distributed fraction of dark matter became much less than unity, and remained that way in spite of subsequent tidal stripping.

\subsection{The Collision Location}\label{sub:CollisionLocation}

It is also worth commenting on our assumption that the relevant point at which to evaluate the velocity of the B cluster is at the scale radius of the A cluster. This is, of course, a simplification: hydrodynamic simulations of the Bullet Cluster collision show significant dynamics even at larger separation radii \cite{BCCollisionNotUnexpected}. Nonetheless, one generically expects that the majority of the collision dynamics will occur at radii not much larger than the scale radius. The exact collision radius chosen should therefore not affect our conclusions by more than order-one factors. To illustrate this, Fig. \ref{fig:AttractivePlot} also shows a conservative version of our bound that instead assumes a collision at two times the scale radius of the A cluster.

\subsection{Dynamical Friction}\label{sub:DynamicalFriction}

Self-interactions of dark matter also lead to a competing effect on the cluster infall velocity: the tendency of scattering to slow the two clusters' relative velocities through dynamical friction, even for purely attractive forces. Dynamical friction is the process through which an object moving through a field of particles with which it interacts gradually transfers kinetic energy to that background field, being slowed in the process \cite{DFOriginal1, DFOriginal2, DFOriginal3, BTGalacicDynamics}.

Dynamical friction tends to slow the passage of cluster B through cluster A, counteracting the additional infall velocity acquired due to the attractive force as discussed above. The magnitude of this effect can be estimated from Chandrashekhar's dynamical friction formula \cite{DFOriginal1, BTGalacicDynamics},
\begin{align}
    \frac{\dd \mathbf{v}_B}{\dd t} &= -\frac{16\pi^2}{3}\alpha^2 G^2 M_B \rho_A \frac{\mathbf{v}_B}{v_B^3} \ln\Lambda, \label{eq:DynamicalFrictionGeneral}
\end{align}
where $M_B$ and $\mathbf{v}_B$ are the mass and velocity of the B cluster, respectively, $\rho_A$ is the mass density of the A cluster (which we will approximate as a constant, since only particles in cluster A within a range $\lambda$ of cluster B contribute, and we are interested in the case where both $\lambda$ and the scale radius of B are significantly smaller than the scale radius of A), and $\Lambda$ is the ratio between the range of the new force and a short-distance cutoff: either the scale radius of the B cluster or the distance at which a cluster A particle would be deflected by $90^\circ$ by its interaction with the B cluster; whichever is larger. (There is no dependence on the velocity distribution function of the A cluster, since the B cluster is moving at A's escape velocity while particles in A must have smaller velocities.)

Note that $\mathbf{v}_B$ is modified, relative to its gravity-only value, in the presence of a new force. In particular, once the potential at the collision point is dominated by the new force rather than by gravity, $v_B \propto \alpha^{1/2}$. This change in velocity also affects the time taken by the infall, and thus the total amount of dynamical friction. The end result of this is that the fractional velocity loss from dynamical friction scales with $\alpha$ as
\begin{align}\begin{split}
    \frac{1}{v_B} \frac{\dd v_B}{\dd r} &= \frac{1}{v_B^2} \frac{\dd v_B}{\dd t} \\
    &= \frac{1}{v_B^2}\left| -\frac{16\pi^2}{3}\alpha^2 G^2 M_B \rho_A \frac{\mathbf{v}_B}{v_B^3} \ln\Lambda \right| \\
    &\propto \frac{\alpha^2}{v_B^4} \propto \alpha^0,
\end{split}\end{align}
i.e. it is independent of the strength of the new force! It is thus sufficient to confirm that dynamical friction's effects are subdominant at the smallest values of $\alpha$ (for each $\lambda$) for which collision velocity could lead to a constraint on new self-interactions to ensure that it is insignificant at all larger values of $\alpha$ as well.

Integrating Eq. \eqref{eq:DynamicalFrictionGeneral} numerically, we find that dynamical friction would only lead to a velocity loss of order $4\%$ for cluster B passing entirely through cluster A, or less than $1\%$ before reaching cluster A's scale radius. Dynamical friction therefore does not significantly change the constraint on new attractive forces obtained above. 

This is also why we do not present a constraint on new forces arising from long-range scattering of dark matter particles in the Bullet Cluster: dynamical friction is precisely the effect of (soft) scattering in this long-range, classical regime, and one might hope to obtain a bound from the fact that it could not have been strong enough to slow the dark matter by order unity. The small effect of dynamical friction compared to the direct effect of the additional attractive potential, however, means that the net effect of the new force is an acceleration, not a deceleration, and thus such a bound would not be physical. While both effects would tend to slow the dark matter in the case of a repulsive potential, the smallness of dynamical friction would still not significantly affect the collision velocity bound obtained above, and---as we will see in Section \ref{sub:HaloBinding}---this is not competitive with bounds on repulsive forces obtained from ultrafaint dwarf galaxies.

\subsection{Tidal Distortion}\label{sub:TidalDistortion}

The discussion so far has ignored any internal dynamics of the two clusters within the Bullet Cluster. However, deformation of the dark matter profiles of the two clusters could lead to a change in the acceleration of the clusters' gas contents, modifying the resulting collision velocity.

As we argued in Section \ref{sub:Separation}, the source of net new forces on cluster B is the gradient of the dark matter density of cluster A, leading to a parametric suppression of forces of range $\lambda < r^*_A$ by a factor of $\lambda/r^*_A$ from the fractional change in density within a distance $\lambda$ of the particle being accelerated. The variation of this force over the size of cluster B is then suppressed by an additional factor of $r^*_B/r^*_A \sim 0.1$. Since our bound corresponds to the new force imparting an additional velocity to the dark matter in cluster B of order the escape velocity from $r^*_A$ of cluster A, the largest possible result of this would be a spread in velocities imparted to opposite sides of cluster B of order $0.1$ times the escape velocity of cluster A, which is in turn smaller than the escape velocity of cluster B by a factor of approximately $4$ due to the higher density of cluster B (see Eq. \eqref{eq:NewPotentialScaleRadiusGeneric}). Cluster B's dark matter profile is thus disrupted by much less than order unity at our bound. Moreover, since the quantities discussed above have identical scaling with $\alpha$ above our bound (when the escape velocity is dominated by the new force rather than by gravity), this remains the case even for larger $\alpha$, ensuring that tidal deformation is never significant.

The dark matter in cluster A is similarly accelerated, and then also decelerated, as it passes through cluster B, leading (in the absence of a restoring force) to a displacement of order $\Delta r \sim \Delta v (r_B/v_B) \sim r_B$ at our bound. This would not be a large effect even if it affected all of cluster A (since $r_B \sim 0.1~r_A$), and additionally it affects only the fraction $(r_B/r_A)^2 \sim 0.01$ of cluster A that passes through cluster B, so the change in the gravitational potential seen by the gas in cluster A is negligible. There is thus no significant change in the A cluster gas's velocity compared to what would have occurred in the presence of gravity alone.

\section{Other Constraints}\label{sec:OtherConstraints}

The Bullet Cluster is far from the only constraint on long-range self-interactions of dark matter. We now turn to some of the other constraints on this parameter space, emphasizing the same key idea presented in Section \ref{sec:CollisionVelocity}: that the gradient of the dark matter density in galaxy and cluster halos causes bounds on new dark matter self-interactions to persist to ranges shorter than the structure length scale, with only a power-law (rather than an exponential) suppression. We will focus on two specific constraints: halos remaining bound even in the presence of new repulsive forces, and the agreement between the cluster masses obtained through gravitational lensing and through measurements of galactic velocities. The same principles should apply to various other constraints on this parameter space \cite{VariousLRDMSIConstraints}, including the mass-to-luminosity ratio of galaxies in the Local Group, tidal stripping of galaxies within clusters, and the modification of rotation curves resulting from galaxy formation. Conversely, see Refs. \cite{DarkForceBackgroundEffects1, DarkForceBackgroundEffects2, DarkForceBackgroundEffects3} for discussion of a bound on attractive forces with long ranges---tentatively extended to $\lambda \gtrsim 40$ kpc---based on the cosmological effects of a background field of the force mediator, and in particular its tendency to modify large-scale structure. It remains unknown, however, how this bound scales at shorter ranges. Similarly, Refs. \cite{TidalTails1, TidalTails2} showed that forces with range $\lambda \gtrsim 10~{\rm kpc}$ and strength $\alpha \gtrsim 0.2$ would result in asymmetries in the tidal tails of the Sagittarius (Sgr) Dwarf Spheroidal Galaxy inconsistent with observations---although Ref. \cite{TidalTailsUpperBound} demonstrates that this effect disappears once $\alpha \sim 1$, and an exact upper bound to this constraint has yet to be determined---but it is unclear how this constraint extends to shorter force ranges, due to the large effects of such a new force on the dark matter profile in the vicinity of the Sgr dwarf; see Section \ref{sub:GravitationalLensing}. Other constraints on long-range self-interactions of dark matter are discussed in Refs. \cite{DMSIMergers, DMSIStructureFormation, DMSIVarious, DarkPlasmasBC}.

\begin{figure}[t]
    \centering
    \includegraphics[width=0.7\linewidth]{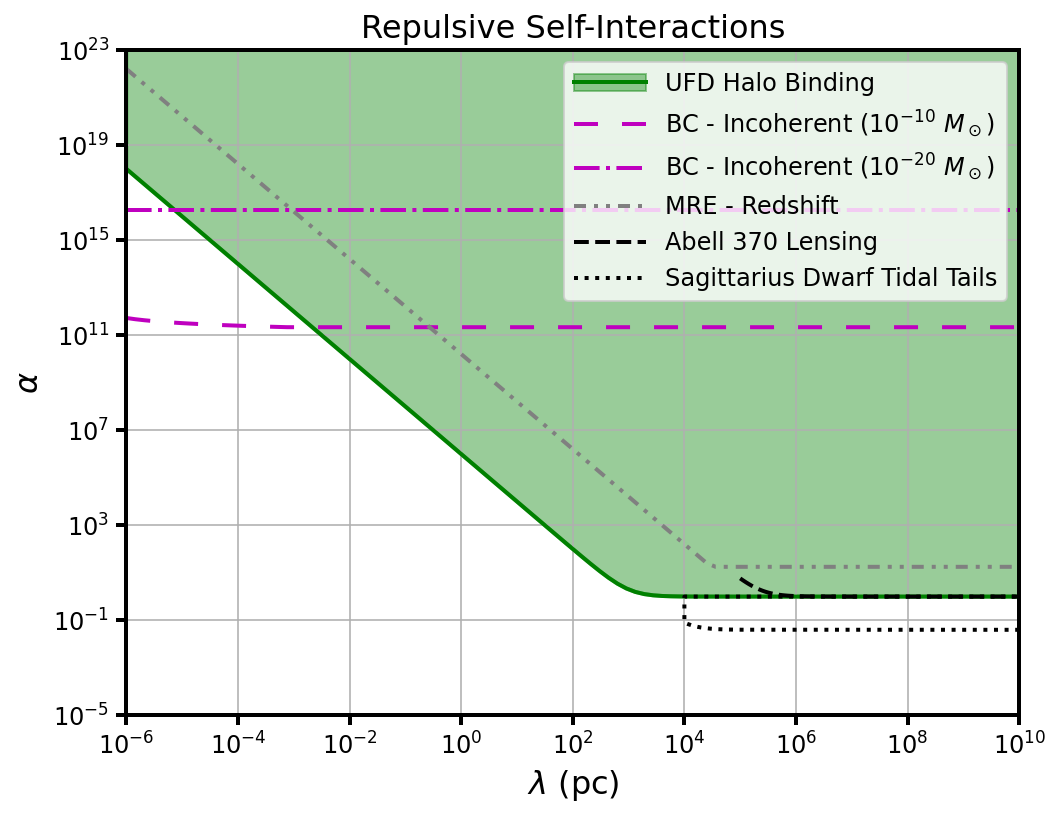}
    \caption{Our constraints on repulsive self-interactions of dark matter from self-binding of ultrafaint dwarf galaxies (see Section \ref{sub:HaloBinding}) and from mass measurements of the Abell 370 cluster (see Section \ref{sub:GravitationalLensing}), and, for comparison, previous constraints from the tidal streams of the Sagittarius dwarf galaxy \cite{TidalTails1, TidalTails2, TidalTailsUpperBound} and from incoherent scattering in the Bullet Cluster \cite{BCFirstDMSILimits, LDMReviewKLZ, BCDMSILimits2, BCDMSILimits3} at two assumed dark matter particle masses (the largest mass at which promordial black holes can make up all of dark matter, $\sim 10^{-10}~M_\odot$ \cite{PBHDMReview}, as well as $\sim 10^{-20}~M_\odot$ to illustrate the mass dependence). The shortest ranges at which we plot the lensing and tidal tails constraints are approximate; see the discussion in Section \ref{sec:OtherConstraints}. Note also that the upper edge of the parameter space excluded by the Sagittarius Dwarf is presently unknown; Ref. \cite{TidalTailsUpperBound} has shown only that $\alpha=1$ is consistent with observation.}
    \label{fig:RepulsivePlot}
\end{figure}

\subsection{Halo Binding}\label{sub:HaloBinding}

A very straightforward bound on repulsive forces arises simply from the existence of dark matter halos, which would unbind in the presence of a sufficiently strong repulsive self-interaction \cite{VariousLRDMSIConstraints}. The smallest observed dark matter halos are the ultrafaint dwarf galaxies' halos with scale radii of approximately $1~{\rm kpc}$ \cite{UFDData, UFDHaloShape}. Repulsive forces with ranges longer than this length scale must therefore be weaker than gravity.

The corresponding bound on shorter-range forces is, as in the Bullet Cluster case, suppressed by a factor of $(1~{\rm kpc}/\lambda)^2$, as we can see from requiring that the total acceleration at the scale radius is still pointing radially inward:
\begin{align}
    a_{\rm DM}(r^*) + a_{\rm g}(r^*) < 0.
\end{align}
These are given by the $a_{\rm DM}$ and $a_g$ in Eqs. \eqref{eq:NewAccelerationDM} and \eqref{eq:GravityAccelerationScaleRadius} but now for a single halo with some density $\rho$ and size $r^* \sim {\rm kpc}$. The resulting constraint is then approximately
\begin{align}
    \alpha \lesssim 1 + \left(\frac{\rm kpc}{\lambda}\right)^2,
\end{align}
where the constant value of $1$ accounts for the large-$\lambda$ behavior discussed below Eq. \eqref{eq:NewAccelerationDM} and we have omitted order-one numerical factors since these depend on the essentially unknown halo profiles of ultrafaint dwarf galaxies. We plot this constraint in Fig. \ref{fig:RepulsivePlot} as ``UFD Halo Binding.''

A closely related constraint on attractive forces arises from the need for orbiting dark matter to be able to stay in orbit, rather than to collapse into a black hole (i.e. for the necessary orbital velocity of the dark matter to be $<1$). It is straightforward to see that this leads to bounds significantly weaker than those calculated in Section \ref{sec:CollisionVelocity}, however.

\subsection{Gravitational Lensing}\label{sub:GravitationalLensing}

We next consider an example of a constraint on dark matter self-interactions that does not obey the power-law behavior discussed thus far at sufficiently small length scales due to qualitatively new dynamics in the short-range, strong-coupling regime. In the very long-range regime, constraints on dark matter self-interactions have been derived from the agreement of cluster mass measurements using the velocities of galaxies (which would be affected by a new force) and using gravitational lensing of light (which would not) \cite{VariousLRDMSIConstraints}. We now consider the short-range behavior of this constraint.

For force ranges $\lambda$ much shorter than the scale radius $r^* \sim 1.2~{\rm Mpc}$, the additional acceleration of dark matter at a radius $r \sim 220 \, {\rm kpc}$ with $\lambda \ll r \ll r^*$ can naively be obtained similarly to Eq. \eqref{eq:NewAccelerationDM}, but now accounting for the faster variation of density near the center of NFW halos:
\begin{align}
    |a_{\rm DM}(r)| \sim \alpha G \frac{\lambda^3}{\lambda^2} \left(\lambda \frac{\dd}{\dd r}\rho_{\rm NFW}(r)\right) \sim \alpha G \frac{\rho_{\rm NFW}(r) \lambda^2}{r}. \label{eq:NFWLensingAcceleration}
\end{align}
Here we have dropped various order-one factors since this is intended only as a rough estimate of the result. For this change in acceleration be no greater than the gravity-only prediction, i.e.~for the velocity distribution (and thus the mass estimate) to not be modified by more than order one, requires that
\begin{align}
    |a_{\rm DM}(r)| \lesssim |a_g(r)| \sim G\rho(r)r,
\end{align}
which leads to a constraint on $\alpha$; this is plotted in Figs. \ref{fig:AttractivePlot} and \ref{fig:RepulsivePlot} as ``Abell 370 Lensing.'' 

However, once $\lambda \lesssim 100$ kpc, the typical size of individual galaxy halos, a limit cannot be derived as explained above because we can no longer ignore the back-reaction of the galaxy halo on to the background dark matter density: the galaxy halo is a significant over-dense region compared to the ambient background density and hence dominates the total dark matter mass contained within $\lambda$ of a typical galaxy halo. There is therefore significant back-reaction on the cluster profile due to the presence of the galaxy. While this does not preclude the possibility of   some limit existing, a detailed simulation would be required to evaluate this back-reaction and estimate the galaxy velocity dispersion in presence of a new short-ranged force stronger than gravity. For this reason, we truncate the limit in Figs. \ref{fig:AttractivePlot} and \ref{fig:RepulsivePlot} at $100$ kpc.

It is important to point out that this back-reaction effect is unique to this case: both the Bullet Cluster interactions considered in Sec.~\ref{sec:CollisionVelocity} and the halo binding discussed in Sec.~\ref{sub:HaloBinding} were based on the smooth components of those objects, which lack significant over-densities.

\section{Conclusion}\label{sec:Conclusion}

Many previous works have presented constraints on interactions between dark matter particles, but they have generally focused on one of two regimes: short-range interactions which result in scattering between individual particles, and long-range (or infinite-range) interactions which effectively modify the strength of gravity within dark sectors. In this work, we have considered the intermediate regime, in which dark matter particles feel a force with range between the dark matter's interparticle spacing and the length scale of the dark matter halos.

If dark matter halos had sharp cutoffs---that is, if they had approximately constant density within some region and negligible density outside of it---the net force resulting from such a new interaction would be exponentially suppressed at distances from the halo greater than the interaction's range. Previous works have therefore occasionally assumed that meaningful constraints cannot be obtained from astrophysical observations for such short-range forces.
In this work, we have shown that this is not the case: the gradual variation in the dark matter density of observed (and simulated) dark matter halos leads only to a power-law suppression in the effects of a finite-range interaction, as a dark matter particle within such a halo sees a net force resulting from the difference in the dark matter density between the regions within $\lambda$ of it that are closer and farther from the halo center. Since halo profiles are believed to vary with radius as power laws, this density contrast is also suppressed only as a power law, and thus finite-range interactions can still have significant collective effects at ranges shorter than the halo length scale.

We then used this behavior to extend various constraints on (effectively) infinite-range interactions to finite ranges; the resulting constraints are plotted in Figs. \ref{fig:AttractivePlot} and \ref{fig:RepulsivePlot}. For illustrative purposes, we focused on the effects of such a finite-range attractive force on the Bullet Cluster, where it results in an acceleration of the smaller cluster in that collision in a way that would result in either a collision velocity of the clusters' gas inconsistent with observations, or with a separation between the dark matter and galaxies within each cluster. For repulsive forces, we noted that a strong constraint can be obtained simply from the fact that ultrafaint dwarf galaxies' halos remain bound. The resulting bounds are the strongest existing bounds on net-charged dark matter self-interactions over at least five orders of magnitude in force range even for dark matter at the heaviest masses allowed for primordial black hole dark matter, and over far greater ranges for typical particle-mass dark matter.

As we have argued, this finite-range behavior is quite generic, with caveats arising primarily from the unknown substructure of dark matter halos. The results of such substructure do not change the two constraints mentioned above by more than order one unless almost all dark matter is concentrated in widely-separated substructures. Other constraints, however, can be much more sensitive to substructure effects: for example, as we showed, constraints from the agreement of galaxy cluster mass measurements between velocity-based and gravitational-lensing based approaches become ineffective once the majority of the dark matter mass within $\lambda$ of a galaxy is bound to that galaxy already, as the galaxy begins to have a large impact on the distribution of dark matter in its vicinity. Even in this case, however, it may be possible to set constraints on dark matter self-interactions via N-body simulations of galactic dynamics; we leave this to future work.

More generally, there are many ways to both refine the constraints we present and extend others to finite ranges. The focus of this work has been the parametric behavior of various present-day constraints at finite ranges, so we have computed those constraints only up to order-one factors. It may be possible to strengthen them somewhat, however, by performing N-body simulations of the bullet cluster collision, galaxy cluster dynamics, or similar. Moreover, there are many other constraints which have been computed at infinite ranges but which we have not considered here. In particular, the existence of a new finite-range dark matter self-interaction stronger than gravity would likely significantly modify structure formation; we leave the study of such cosmological effects to future work.

\appendix
\section{Finite Density Effects}\label{app:FiniteDensityEffects}

In most of this work, we have remained largely agnostic to the specific dark matter model leading to the assumed Yukawa force, since the effects we considered are fairly model-independent. Introducing a long-range interaction to a dark matter model may have other, more non-trivial effects, however. In particular, a finite background density of dark matter can modify the mass, expectation value, and other properties of the new force mediator, significantly changing its behavior compared to the zero-density limit.

\subsection{Scalar Mediators - Small Couplings}\label{subapp:ScalarMediators}

Since the focus of this paper was on always-attractive interactions, we begin by considering a model of fermionic dark matter $\chi$ with a scalar mediator $\phi$:
\begin{align}\begin{split}
    \mathcal{L} \supset \;& \overline{\chi}(i\cancel\partial - m)\chi + \frac{1}{2}(\partial_\mu \phi)^2 - \frac{1}{2}\mu^2 \phi^2 \\
    &+ g\phi\overline{\chi}\chi + \lambda_3 \phi^3 + \lambda_4 \phi^4.
\end{split}\end{align}
Most of this work parametrized the new force interaction strength via $\alpha$, the strength relative to gravity instead; this is related to $g$ by
\begin{align}
    g^2 = 4\pi \alpha G m^2.
\end{align}

Note first that loops of $\chi$ will induce potentially large values for $\mu$, $\lambda_3$, and $\lambda_4$:
\begin{subequations}\begin{align}
    (\Delta \mu^2)_{\rm loop} &\sim \frac{g^2\Lambda^2}{16\pi^2} \\
    (\Delta \lambda_3)_{\rm loop} &\sim \frac{g^3\Lambda}{16\pi^2} \\
    (\Delta \lambda_4)_{\rm loop} &\sim \frac{g^4}{16\pi^2}
\end{align}\end{subequations}
with $\Lambda$ an ultraviolet scale. Even for $\Lambda \sim {\rm GeV}$ and $\alpha=1$, this contribution to $\mu^2$ is of order $10^{-11}$ eV, corresponding to a force range of order $10^{-13}$ pc. Since we are primarily interested in forces with much larger ranges than this, the models considered in this work will generally be quite fine-tuned. Similarly, we have ignored the effects of any $\lambda_3$ or $\lambda_4$ in this work for simplicity, although this too would likely require fine tuning.

Additional contributions to these terms arise within a dark matter background, however, due to the non-zero expectation value of $\overline{\chi}\chi$. Unlike the loop corrections discussed above, these terms cannot be generally fine-tuned away since they are functions of the (variable) local dark matter density. We therefore now turn to evaluating these contributions and determining whether they modify the results in the main text.

In a background of $\chi$ number density $n_\chi$, we have $\left\langle \overline\chi\chi \right\rangle = n_\chi$ and thus
\begin{align}
    V'(\phi) = \frac{1}{2}\mu^2\phi^2 - g n_\chi \phi
\end{align}
(assuming, as before, that $\lambda_3=0$ and $\lambda_4=0$). Note that we use the number density of $\chi$ rather than the dark matter density since the latter may include contributions from $\phi$. (We will return to the latter possibility in the next section; numerical values in this section assume that the dark matter density is given by $m n_\chi$, although algebraic expressions are left general.) Minimizing this potential gives an expectation value for $\phi$ of
\begin{align}
    \expect{\phi} = \frac{g n_\chi}{\mu^2} = \sqrt{4\pi\alpha G}\frac{mn_\chi}{\mu^2}.
\end{align}
Evaluating this expectation value at the scale radius of the B cluster, where $\rho_B(r_B^*) \sim 9\times10^{-4} ~M_\odot\,{\rm pc}^{-3}$, and assuming that all of the dark matter density comes from $\chi$ (i.e. that $\rho_\DM = mn_\chi$) gives
\begin{align}
    \expect{\phi}_{r_B^*} \sim 2\times10^{24}~{\rm eV}~\sqrt{\alpha}\pfrac{\lambda}{\rm Mpc}^2.
\end{align}
This is, roughly, the largest density we are concerned with in this work, so we will use it as a conservative benchmark in the rest of this appendix.

The additional energy density resulting from this background value of $\phi$ is, at zero temperature,
\begin{align}
    \Delta\rho = \frac{1}{2}\mu^2\expect{\phi}^2 - gn_\chi\expect{\phi} = -2\pi\alpha G\frac{m^2 n_\chi^2}{\mu^2}. \label{eq:DensityChangeNR}
\end{align}
Evaluating this at the same benchmark density of $\rho_B(r_B^*)$ gives
\begin{align}
    \Delta\rho \sim -2\times10^{-7}\frac{M_\odot}{{\rm pc}^3}~\alpha\pfrac{\lambda}{\rm Mpc}^2,
\end{align}
a subdominant density contribution at all points along our bound with $\lambda \ll 20$ Mpc, but not at larger $\lambda$ or for $\alpha$ more than three orders of magnitude above our bound. We consider the effects of larger values of $\lambda$ and $\alpha$ below.

The expectation value for $\phi$ also shifts the mass of $\chi$ by
\begin{align}
    \Delta m = -g\expect{\phi} = -\frac{g^2 n_\chi}{\mu^2} = -4\pi\alpha G \frac{m^2 n_\chi}{\mu^2},
\end{align}
or, in a more convenient form, by a relative amount
\begin{align}
    \frac{\Delta m}{m} = -4\pi\alpha G\frac{m n_\chi}{\mu^2}.
\end{align}
Note again that this result is correct only when $m n_\chi$ makes up all of the dark matter density. Evaluating the change in the mass of $\chi$ at the scale radius of B, we have
\begin{align}
    \frac{\Delta m}{m} \sim 5\times10^{-4}~ \alpha\pfrac{\lambda}{\rm Mpc}^2.
\end{align}
The $\Delta m/m \ll 1$ condition is thus satisfied in essentially the same parameter space as the $\Delta\rho \ll \rho_\DM$ condition discuss previously.

The introduction of a new attractive force also leads to a new Jeans scale, which is given by
\begin{align}
    \lambda_J \sim c_s t_{\rm ff} \sim \sqrt{\frac{P/\rho_\DM}{\alpha G\rho_\DM}} \sim \sqrt{\frac{v_\DM^2}{\alpha G\rho_\DM}} \sim m_{\rm Debye}^{-1}
\end{align}
where $c_s$ is the speed of sound, $t_{\rm ff}$ is the characteristic free-fall time at the given density and coupling, $P$ is the pressure of the dark matter, and we have assumed that $\lambda_J < \lambda$ such that the new force is essentially long-ranged. At longer length scales, $\alpha$ is effectively suppressed by the usual quadratic factor of $(\lambda_J/\lambda)^2$; it is straightforward to confirm that this does not lead to a new Jeans instability, so either $\lambda_J \lesssim \lambda$ or there is no Jeans scale from the new force at all.

Evaluating the Jeans length (in the long-$\lambda$ regime) at our benchmark density gives
\begin{align}
    \lambda_J \sim 2~{\rm Mpc}~\frac{1}{\sqrt{\alpha}} \pfrac{v_\DM}{4000~{\rm km/s}}
\end{align}
which is right around the edge of our excluded region below $\lambda \sim 1$ Mpc, but is below it for larger ranges. This should not affect our conclusions, however: while there may be additional constraints associated with the new Jeans instability, all of the arguments in the main text hold regardless.

There are also background contributions to $\lambda_3$ and $\lambda_4$:
\begin{align}
    \Delta\lambda_3 &\sim \frac{g^3\rho_\DM}{m^3} \sim 6\times10^{-90}~\alpha^{3/2}~{\rm eV} \\
    \Delta\lambda_4 &\sim \frac{g^4\rho_\DM}{m^4} \sim 2\times10^{-117}~\alpha^2.
\end{align}
Here we have again shown the values at our benchmark density. These values are small enough that they should have no practical significance for any parameters of interest.

\begin{figure}[b]
    \centering
    \includegraphics[width=0.7\linewidth]{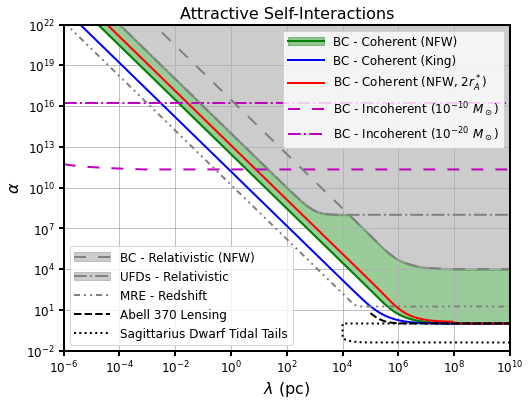}
    \caption{Our constraints on attractive self-interactions of dark matter beyond gravity, extended from Fig. \ref{fig:AttractivePlot}. Three new lines have been added here: In grey, we show the regions where dark matter would need to be relativistic in order to achieve the observed density in the B cluster and in typical UFDs; this is also the regime in which the background density of the force mediator becomes comparable to that of the dark matter, as described in Appendix \ref{subapp:LargeCouplings}. The dash-dot-dotted line shows where the latter effect occurs at matter-radiation equality, which may or may not lead to observable effects on structure formation; see, again, Appendix \ref{subapp:LargeCouplings}.}
    \label{fig:AttractivePlotExtended}
\end{figure}

\subsection{Scalar Mediators - Large Couplings}\label{subapp:LargeCouplings}

In the previous section, our phenomenological discussion focused on small couplings, for which the density arising from $\expect{\phi}$ is negligible and we could use $\rho_\DM = mn_\chi$. In this section we consider the larger-coupling regime in which this is no longer the case. 

We can generalize our result \eqref{eq:DensityChangeNR} from the previous section for the change in density from the expectation value of $\phi$ to situations where $\chi$ is relativistic as
\begin{align}
    \Delta\rho = \frac{1}{2}\mu^2\expect{\phi}^2 - g\frac{n_\chi}{\gamma}\expect{\phi} = -2\pi\alpha G\frac{m^2 n_\chi^2}{\gamma^2\mu^2}, \label{eq:BackgroundDensity}
\end{align}
where we have accounted for $n_\chi$ (but not $\bar{\chi}\chi$) being relativistically enhanced by the Lorentz factor $\gamma$. Note that this result is negative, corresponding to the binding energy from the new attractive force. This will be reduced after accounting for kinetic energy---by a factor of 2, in the long-range limit, assuming virialization---but should not generally be parametrically reduced. 

This negative contribution to the density is quadratic in $n_\chi$, and thus sets a maximum density $m n_\chi\gamma+\Delta\rho$ that could be achieved simply by packing in more and more $\chi$ particles of a particular energy. Higher densities can still be achieved, however, by making the $\chi$ particles relativistic. Requiring that the total density be positive, i.e. that $|\Delta\rho| \lesssim \rho_\chi = m n_\chi\gamma$ requires, roughly,
\begin{align}
    \gamma^4 \gtrsim \alpha G\frac{mn_\chi\gamma}{\mu^2} \geq \alpha G\frac{\rho_{\rm obs}}{\mu^2} \label{eq:OrderOneEffectCondition}
\end{align}
where $\rho_{\rm obs} \leq mn_\chi\gamma$ is the observed dark matter density of the halo in question. This sets a minimum value for the typical Lorentz factor in a halo of
\begin{align}
    \gamma_{\rm min} \sim \pfrac{\alpha G\rho_{\rm obs}}{\mu^2}^{1/4}.
\end{align}
The resulting object, for $\gamma_{\rm min} > 1$, is bizarre---a dark matter halo of highly relativistic particles, in which the kinetic energy and the binding energy almost exactly cancel to give a small total mass---but there is no immediate contradiction. We therefore want to confirm that our bounds will still extend to this high-$\alpha$ regime.

First, note that we can also set a maximum typical Lorentz factor $\gamma_{\rm max}$ from the fact that $\rho_\chi + \Delta\rho$ should always be less than $mn_\chi$ in order for the halo to be stable; if this were not the case, it would be energetically favorable for the $\chi$ particles to disperse over a large volume (i.e. a preference for low number density at fixed total particle number). It is easy to confirm that $\rho_{\rm tot}(\gamma_{\rm min}+1) \geq mn_\chi$ and thus $\gamma_{\rm max} \leq \gamma_{\rm min}+1$; this will be sufficient for our purposes.

Combining our two inequalities, we can now calculate the $\gamma$ required to explain an observed halo's density as a function of $\alpha$, up to a small uncertainty:
\begin{align}
    \gamma(\rho_{\rm obs},\alpha) \in \left[\gamma_{\rm min},\gamma_{\rm min}+1\right]
\end{align}
and thus, for $\gamma_{\rm min} \gg 1$,
\begin{align}
    \gamma(\rho_{\rm obs},\alpha) \sim \pfrac{\alpha G\rho_{\rm obs}}{\mu^2}^{1/4} \propto \alpha^{1/4}.
\end{align}

This is, finally, sufficient to confirm that the bounds in this work will still apply to this high-$\alpha$ regime. Recall that this work's bound is, up to order-one factors, equivalent to requiring that the depth of the potential well seen by the B cluster not be changed by more than order-one by the presence of a new force. For halos of relativistic particles, the total charge of each halo (at a fixed mass) is suppressed relative to the non-relativistic case by a factor of $\gamma$ (since the energy per particle, but not the charge per particle, is enhanced by this factor). The new force potential thus scales as
\begin{align}
    V_{\rm new} \propto \frac{\alpha}{\gamma^2} \propto \alpha^{1/2}.
\end{align}
Crucially, this potential continues to grow with $\alpha$, in spite of the relativistic suppression. Values of $\alpha$ larger than the minimum values that we exclude in this work are thus still excluded, in spite of this effect. While this argument precludes these types of relativistic dark matter halos from appearing in nature, we show the parameter space for which they would have been present in the bullet cluster (assuming an NFW profile) and in typical ultrafain dwarf galaxies as ``BC - Relativistic (NFW)'' and ``UFDs - Relativistic,'' respectively, in Fig.~\ref{fig:AttractivePlotExtended}.

More generally, couplings well above the limits in this work may significantly modify halo profiles and structure formation, but this should not change our conclusions.

There may, however, also be effects on structure formation at couplings below those we exclude, potentially leading to stronger constraints. In particular, Eq. \eqref{eq:OrderOneEffectCondition} may not be satisfied in the early universe, although effects of the new force will generally be suppressed by the near-homogeneity of the universe at this time. Such effects may (or may not) change dark matter dynamics in observable ways; we leave exploration of this to future work. For reference, we show the couplings required to saturate Eq. \eqref{eq:OrderOneEffectCondition} at matter-radiation equality, with the force range cut off by the inverse Hubble scale, in Fig. \ref{fig:AttractivePlotExtended} as ``MRE - Redshift.''

\subsection{Vector Mediators}\label{subapp:VectorMediators}

If our fermionic dark matter interacts instead via a vector mediator, we have
\begin{align}\begin{split}
    \mathcal{L} \supset \;& \overline{\chi}(i\cancel\partial - m)\chi - \frac{1}{4}G_{\mu\nu}G^{\mu\nu} - \mu^2A_\mu A^\mu \\
    &- g\overline{\chi}\gamma^\mu\chi A_\mu
\end{split}\end{align}
where $A_\mu$ is the new vector field and $G_{\mu\nu} = \partial_\mu A_\nu - \partial_\nu A_\mu$. (We will remain agnostic to the UV theory leading to this Lagrangian at the low energies we are interested in.)

In this case, background expectation values of $A_\mu$ arise from $\expect{\overline{\chi}\gamma^\mu\chi}$ rather than simply $n_\chi$, corresponding to the potential for net-neutral plasmas. Assuming, as we do in the main text, that all of the dark matter has the same sign of charge, however, we have charge density $J_0 = \expect{\overline{\chi}\gamma^0\chi}$ and current density $J_i \sim \expect{\overline{\chi}\gamma^i\chi} \sim J_0\expect{v_i}$ with $\expect{v_i}$ the average dark matter velocity at a given location; we expect the latter to be subdominant, since dark matter should be non-relativistic.

Many of the effects resulting from this background density are similar to the scalar case, so we do not repeat their estimation here. One crucial difference, however, is that the density contribution from the mediator is now positive rather than negative. (Recall that the vector mediator case corresponds to a repulsive force, whereas scalar mediators lead to attractive forces.) Since this density is quadratic in the number density of $\chi$---see Eq. \eqref{eq:BackgroundDensity}---this density contribution will redshift as kination rather than matter; it must therefore be a subdominant density contribution at all times after matter-radiation equality. This does not lead to interesting new constraints, however, since this occurs at couplings strictly larger than those we have already excluded in this work, even before accounting for the small density fluctuations in the early universe (compare the matter-radiation equality line of Fig. \ref{fig:AttractivePlotExtended} with our exclusion plot for repulsive forces, Fig. \ref{fig:RepulsivePlot}).

It is also worth noting that Debye screening, while typically present in plasmas, does not affect our conclusions in this scenario: Debye screening is the result of a rearrangement of charges to screen the effect of a test charge, but here we are interested in the effects of the charge distribution of the halo as a whole, which cannot be screened. (Put another way, our conclusions have already accounted for any rearrangement of charges, as we take the density profile of the halo as an input.) While the introduction of a new force can certainly modify halo profiles, our results are largely agnostic to the details of the halo shape, as we discuss in the main text. Debye screening therefore does not affect the bounds in the main text, although it may be relevant for other physical processes.

\acknowledgments

The authors acknowledge support by NSF Grants PHY-2310429 and PHY-2014215, Simons Investigator Award No.~824870, DOE HEP QuantISED award \#100495, the Gordon and Betty Moore Foundation Grant GBMF7946, and the U.S.~Department of Energy (DOE), Office of Science, National Quantum Information Science Research Centers, Superconducting Quantum Materials and Systems Center (SQMS) under contract No.~DEAC02-07CH11359.
ZB is supported by the National
Science Foundation Graduate Research Fellowship under Grant No.~DGE-1656518 and by
the Dr. HaiPing and Jianmei Jin Fellowship
from the Stanford Graduate Fellowship Program.


\bibliographystyle{JHEP}
\bibliography{biblio.bib}

\end{document}